      [1999/12/01 v1.4c Il Nuovo Cimento]
\documentclass{cimento}

\newcommand{\be}{\begin{eqnarray}}
\newcommand{\ee}{\end{eqnarray}}
\newcommand{\bea}{\begin{eqnarray}}
\newcommand{\eea}{\end{eqnarray}}

\title{The Nucleon Spin Sum Rule}
\author{M.~Burkardt\from{ins:x} }
\instlist{\inst{ins:x} New Mexico State University - Las Cruces, N.M.  }
\PACSes{\PACSit{14.20}{Dh}}
\begin{document}

\maketitle

\begin{abstract}
Definitions of orbital angular momentum based on Wigner distributions
are used as a framework to discuss the connection between the Ji definition
of the quark orbital angular momentum and that of Jaffe and Manohar.
We find that the difference between these two definitions can be interpreted as the change in the quark orbital angular momentum due to final state interactions as it leaves the target in a DIS experiment.
\end{abstract}

\section{Angular Momentum Decompositions}

Since the famous EMC experiments revealed that only a small fraction
of the nucleon spin is due to quark spins \cite{EMC}, 
there has been a great
interest in `solving the spin puzzle', i.e. in decomposing the
nucleon spin into contributions from quark/gluon spin and
orbital degrees of freedom.
In this effort, the Ji decomposition \cite{JiPRL}
\begin{equation}
\frac{1}{2}=\frac{1}{2}\sum_q\Delta q + \sum_q { L}_q^z+
J_g^z
\label{eq:JJi}
\end{equation}
appears to be very useful: through GPDs,
not only the quark spin contributions $\Delta q$ but also
the quark total angular momenta $J_q \equiv \frac{1}{2}\Delta q + 
{ L}_q^z$ (and by subtracting the spin piece also the
the quark orbital angular momenta $L_q^z$) entering this decomposition
can be accessed experimentally.
The terms in (\ref{eq:JJi}) are defined as expectation
values of the corresponding terms in the angular momentum tensor
\begin{equation}
M^{0xy}= \sum_q \frac{1}{2}q^\dagger \Sigma^zq +
\sum_q q^\dagger \left({\vec r} \times i{\vec D}
\right)^zq
+  
\left[{\vec r} \times \left({\vec E} \times {\vec B}\right)\right]^z
\label{M012}
\end{equation}
in a nucleon state with zero momentum. Here
$i{\vec D}=i{\vec \partial}-g{\vec A}$ is the gauge-covariant
derivative.
The main advantages of this decomposition are that each term can be 
expressed as the
expectation value of a manifestly gauge invariant
local operator and that the
quark total angular momentum $J^q=\frac{1}{2}\Delta q+L^q$
can be related to GPDs 
\cite{JiPRL} 
and is thus accessible in deeply virtual Compton scattering and
deeply virtual meson production and can also be
calculated in lattice gauge theory. 

Jaffe and Manohar have proposed an alternative decomposition of the
nucleon spin, which does have a partonic interpretation
\cite{JM}, and in which also two terms, $\frac{1}{2}\Delta q$ and $\Delta G$,
are experimentally accessible
\begin{equation}
\frac{1}{2}=\frac{1}{2}\sum_q\Delta q + \sum_q {\cal L}^q+
\Delta G + {\cal L}^g.
\label{eq:JJM}
\end{equation}
The individual terms in (\ref{eq:JJM}) can be defined as matrix elements of the corresponding
terms in the $+12$ component of the angular momentum tensor
\begin{equation}\hspace*{0.3cm}
M^{+12} \!= \frac{1}{2}\sum_q q^\dagger_+ \gamma_5 q_+ +
\sum_q q^\dagger_+\!\!\left({\vec r}\times i{\vec \partial}
\right)^z \!\!\!q_+  
+ \varepsilon^{+-ij}\mbox{Tr}F^{+i}A^j
+ 2 \mbox{Tr} F^{+j}\!\left({\vec r}\times i{\vec \partial} 
\right)^z\!\!\!\! A^j
\label{M+12}
\end{equation}
for a nucleon polarized in the $+\hat{z}$ direction.
The first and third term in (\ref{eq:JJM}),(\ref{M+12}) are the
`intrinsic' contributions (no factor of ${\vec r}\times $) 
to the nucleon's angular momentum $J^z=+\frac{1}{2}$ and have a 
physical interpretation as quark and gluon spin respectively, while
the second and fourth term can be identified with the quark/gluon
OAM.
Here $q_+ \equiv \frac{1}{2} \gamma^-\gamma^+ q$ is the dynamical
component of the quark field operators, and light-cone gauge
$A^+\equiv A^0+A^z=0$ is implied. 
The residual gauge invariance can be fixed by
imposing anti-periodic boundary conditions 
${\vec A}_\perp({\bf x}_\perp,\infty)=-
{\vec A}_\perp({\bf x}_\perp,-\infty)$ on the transverse components
of the vector potential.
${\cal L}$ also naturally arises in a light-cone wave function description of
hadron states, where $\frac{1}{2}=\frac{1}{2}\sum_q \Delta q + \Delta G+
{\cal L}$, in the sense of an eigenvalue equation, is manifestly satisfied for each Fock component individually \cite{LCWF}.

A variation of (\ref{eq:JJi}) has been
suggested in Ref. \cite{Wakamatsu}, where part of $L_q^z$ is attributed to the
glue as 'potential angular momentum'. As we will discuss in the following, the potential angular momentum also has a more physical interpretation as the effect
from final state interactions.
Other decompositions, in which only one
term is experimentally accessible, will not be discussed in this brief note.

\section{Orbital Angular Momentum from Wigner Distributions}
Wigner distributions can be defined as 
defined as off forward matrix elements of non-local
correlation functions \cite{wigner, jifeng,metz}
\begin{equation}\quad
W^{\cal U}(x,{\vec b}_\perp, {\vec k}_\perp)
\equiv \!\int \!\!\frac{d^2{\vec q}_\perp}{(2\pi)^2}
\!\!\int \!\frac{d^2\xi_\perp d\xi^-}{(2\pi)^3}
e^{-i{\vec q}_\perp \cdot {\vec b}_\perp}
e^{i(xP^+\xi^-\!-{\vec k}_\perp\cdot{\vec \xi}_\perp)}
\langle P^\prime S^\prime |
\bar{q}(0)\Gamma {\cal U}_{0\xi}q(\xi)|PS\rangle
\label{eq:wigner}
\end{equation}
with $P^+=P^{+\prime}$, $P_\perp = -P_\perp^\prime = \frac{q_\perp}{2}$.
Throughout this paper, we will chose ${\vec S}={\vec S}^\prime = \hat{\vec z}$. Furthermore, we will focus on the 'good' component by selecting $\Gamma=\gamma^+$.
In order to ensure manifest gauge invariance, a Wilson line gauge link 
${\cal U}_{0\xi}$ connecting the quark field operators at position $0$ and $\xi$ must 
be included \cite{hatta,lorce2}. The issue of choice of path
for the Wilson line will be addressed below. 

In terms  of  Wigner distributions, quark OAM can be defined as \cite{lorce}
\begin{equation}
L_{\cal U}= \int dx d^2{\vec b}_\perp d^2{\vec k}_\perp \left({\vec b}_\perp \times {\vec k}_\perp \right)_z
W^{\cal U}(x,{\vec b}_\perp,{\vec k}_\perp).
\label{def:OAM}
\end{equation}
No issues with Heisenberg's uncertainty principle arise as only perpendicular combinations of position ${\vec b}_\perp$ and momentum ${\vec k}_\perp$ are
needed simultaneously in Eq.(\ref{def:OAM}).

A straight line 
for the Wilson line in ${\cal U}_{0\xi}$ is often the most natural choice, yielding
\cite{jifeng}
\begin{eqnarray} \label{eq:LJi}
L^q_{straight}&\equiv& 
 \int dx d^2{\vec b}_\perp d^2{\vec k}_\perp \left({\vec b}_\perp \times {\vec k}_\perp \right)_z
W^{straight}(x,{\vec b}_\perp,{\vec k}_\perp)\\
&=&
\frac{ \int d^3{\vec r}\langle PS | 
q^\dagger({\vec r}) \left( {\vec r}\times i{\vec D}\right)q({\vec r})|PS\rangle}
{\langle PS |PS\rangle}=L^q_{Ji}\nonumber
\end{eqnarray}
for a nucleon polarized in the $+\hat{z}$ direction,
where
$i{\vec D} = i{\vec \partial} + g{\vec A}({\vec r})$ is the usual gauge-covariant derivative.
This is also the OAM that appears in the Ji-de\-com\-po\-si\-tion (\ref{eq:JJi}).

However, depending on the context, other choices for the path in the Wilson link ${\cal U}$ should be made. Indeed, in the context
of Transverse Momentum dependent parton Distributions (TMDs) probed in Semi-Inclusive Deep-Inelastic Scattering (SIDIS) \cite{ams,RPP,Mauro} the path should be taken to be a straight line to $x^-=\infty$
along (or very close to) the light-cone. This particular choice ensures proper inclusion of the
Final State Interactions (FSI) experienced by the struck quark as it leaves the nucleon
along a nearly light-like trajectory in the Bjorken limit. However, a Wilson line to
$\xi^-=\infty$, for fixed ${\vec \xi}_\perp$ is not yet sufficient to render Wigner distributions
manifestly gauge invariant, but a link at $x^-=\infty$ must be included to ensure manifest
gauge invariance. While the latter may be unimportant in some gauges, it is crucial in
light-cone gauge for the description of TMDs relevant for SIDIS \cite{jifengTMD}. 

\begin{figure}
\unitlength1.cm
\begin{picture}(10,2.6)(1.3,20.2)
\includegraphics{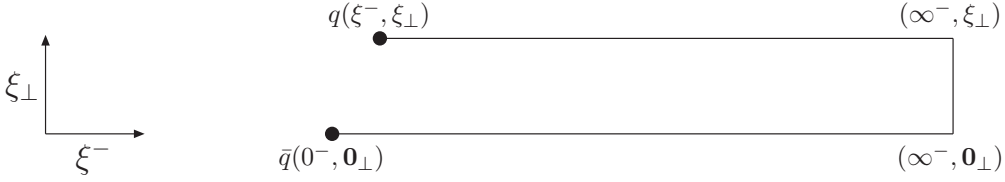}
\end{picture}
\caption{Illustration of the path for the Wilson line gauge link used to define the Wigner distribution $W^{+LC}$ 
(\ref{eq:wigner}).
}
\label{fig:staple}
\end{figure}

Let ${\cal U}^{+LC}_{0\xi}$ be the Wilson path ordered exponential obtained by first taking
a Wilson line from $(0^-,{\vec 0}_\perp)$ to $(\infty,{\vec 0}_\perp)$, 
then to $(\infty,{\vec \xi}_\perp)$, and then to $(\xi^-,{\vec \xi}_\perp)$, with each segment being a straight line (Fig. \ref{fig:staple}) \cite{hatta}. 
The shape of the segment at $\infty$ is irrelevant as the gauge field is pure gauge there, but it is still necessary to include a connection at $\infty$ and for
simplicity we pick a straight line. A a similar 'staple' to $-\infty$ is used to define the Wilson path ordered exponential ${\cal U}^{-LC}_{0\xi}$. Using those light-like gauge links we define
\begin{equation}\quad\quad
W^{\pm LC}(x,{\vec b}_\perp, {\vec k}_\perp)
\equiv \!\!\int\! \frac{d^2{\vec q}_\perp}{(2\pi)^2}\!\!\int \!\frac{d^2\xi_\perp d\xi^-}{(2\pi)^3}
e^{i(xP^+\xi^-\!-{\vec k}_\perp\cdot{\vec \xi}_\perp)}
\langle P^\prime S^\prime |
\bar{q}(0)\Gamma {\cal U}^{\pm LC}_{0\xi}q(\xi)|PS\rangle .
\label{eq:wignerpm}
\end{equation}
This definition for $W^{+LC}$ the same as that in \cite{hatta} and similar to that of $W_{LC}$ in Ref. \cite{jifeng}, except that
the link segment at $x^-=\infty$ was not included in the definition of $W_{LC}$
\cite{jifeng}. The Wilson like gauge link used  to guarantee manifest gauge invariance is defined using a light-like 'staple, i.e. it is constructed using
three straight line gauge links
${\cal U}^{+ LC}_{0\xi} = W_{0^-0_\perp,\infty^-0_\perp}
W_{\infty^-0_\perp,\infty^-\xi_\perp}W_{\infty^-\xi_\perp,\xi^-\xi_\perp}$
and similarly for ${\cal U}^{- LC}_{0\xi}$.

In light-cone gauge $A^+=0$ the Wilson lines to $x^-=\infty$ become trivial and only the
piece at $x^-=\infty$ remains.
While the gauge field at light-cone infinity ${\vec A}_\perp(\pm \infty,{\vec r}_\perp)$ cannot be neglected or set equal to zero in light-cone gauge,
it can be chosen to satisfy anti-symmetric boundary conditions
\begin{equation}
{\vec \alpha}_\perp({\vec r}_\perp)\equiv
{\vec A}_\perp(\pm \infty,{\vec r}_\perp) =-{\vec A}_\perp(\pm \infty,{\vec r}_\perp).
\label{eq:abc}
\end{equation}
This choice maintains manifest PT (sometimes called 'light-cone parity') invariance.

Using these Wigner distributions, one can now proceed to introduce orbital angular momentum
as
\begin{eqnarray}
{\cal L}_{\pm}^q &\equiv& \int dx d^2{\vec b}_\perp d^2{\vec k}_\perp \left({\vec b}_\perp \times {\vec k}_\perp \right)^z
W^{\pm LC}(x,{\vec b}_\perp,{\vec k}_\perp) \label{eq:Lpm}\\
&=& \frac{ \int d^3{\vec r}\langle PS | 
\bar{q}({\vec r}) \gamma^+\left[{\vec r}\times\left( i{\vec \partial}\pm g {\vec \alpha}_\perp ({\vec r}_\perp\right)\right]^zq({\vec r})|PS\rangle}
{\langle PS |PS\rangle}, \nonumber
\end{eqnarray}
and similar for the glue. Eq. (\ref{eq:Lpm}) differs from 
\begin{equation}
{\cal L}^q = \frac{ \int d^3{\vec r}\langle PS | 
\bar{q}({\vec r}) \gamma^+\left({\vec r}\times i{\vec \partial}\right)^zq({\vec r})|PS\rangle}
{\langle PS |PS\rangle}
\end{equation}
(denoted $\tilde{L}^q$ in Ref. \cite{jifeng}) by the contribution from the gauge field $\pm {\vec \alpha}_\perp$ at $\pm \infty$. ${\cal L}^q$ is also identical to the quark OAM appearing in the Jaffe-Manohar
decomposition (\ref{eq:JJM}).
%

\section{Connections Between Different Definitions for OAM}
First of all from  PT invariance one finds that ${\cal L}_+^q={\cal L}_-^q$ \cite{hatta}.
As a corrolary,
since the piece at $\pm\infty$ cancels in the average
both must thus be identical to the OAM appearing in the Jaffe-Manohar decomposition (with antiperiodic boundary conditions for $A_\perp$)
\begin{equation}
{\cal L}^q = \frac{1}{2}\left({\cal L}^q_{+}+{\cal L}^q_{-}\right)={\cal L}_+^q=
{\cal L}_-^q.
\end{equation}
Therefore, even though the gauge link at $x^-=\pm \infty$ is essential for the
description of TMDs \cite{jifengTMD}, it does not contribute to the OAM provided
anti-periodic boundary conditions (\ref{eq:abc}) in light-cone gauge are implied \cite{lorce2}.

To establish the connection with the quark OAM entering the Ji-decomposition,
we consider (for simplicity in light-cone gauge)
\begin{eqnarray}\hspace*{0.3cm}
{\cal L}^q - L^q={\cal L}^q_+ - L^q &=&  \frac{ \int d^3{\vec r}\langle PS | 
\bar{q}({\vec r}) \gamma^+\left[{\vec r}_\perp \times\left( g {\vec \alpha}_\perp ({\vec r}_\perp)
-g{\vec A}_\perp ({\vec r})\right)\right]_zq({\vec r})|PS\rangle}
{\langle PS |PS\rangle}.
\end{eqnarray}
Here we replaced $\gamma^0\rightarrow \gamma^+$ for a nucleon at rest in the definition for $L^q$ \cite{BC}.

Using (in light-cone gauge $A^+=0$ and hence $G^{+i}=\partial_-A^i$)
\begin{equation}
{\alpha}^i ({\vec r}_\perp)
-{A}^i ({\vec r}) = \int_{r^-}^{\infty} dr^- \partial_- {A}^i_\perp ({\vec r}) 
= \int_{r^-}^{\infty} dx^- G^{+i} ({\vec r}) 
\label{eq:integral}
\end{equation}
and noting that
\begin{equation}
T^z({\vec r})\equiv g\left(x G^{+y}({\vec r})-yG^{+x}({\vec r})\right)
\end{equation}
represents the $\hat{z}$ component of the torque that acts on a particle moving with
(nearly) the velocity of light in the $-\hat{z}$ direction -- the direction in which the ejected
quark moves. Thus the difference between the (forward) light-cone  and the
local definitions of the OAM is the
change in OAM due the color force
from the spectators \cite{mb:force} on the active quark
\begin{equation}
{\cal L}^q - L^q = \frac{ \int d^3{\vec r}\langle PS | 
\bar{q}({\vec r}) \gamma^+\int_{r^-}^{\infty}dy^- T^z(y^-,{\vec r}_\perp)
q({\vec r})|PS\rangle} {\langle PS |PS\rangle}.
\label{eq:torque}
\end{equation}
Therefore, while $L^q$ represents the local and manifestly gauge invariant OAM of the
quark {\it before} it has been struck by the $\gamma^*$, ${\cal L}^q$ represents 
the gauge invariant OAM {\it after} it has left the nucleon and moved to $\infty$.
This physical interpretation of the difference between the TMD based (i.e. Jaffe-Manohar) definition of quark OAM with a light-cone staple represents our main result \cite{mb:torque}.

\begin{figure}
\unitlength1.cm
\begin{picture}(10,7.5)(-0.2,13.3)
\includegraphics{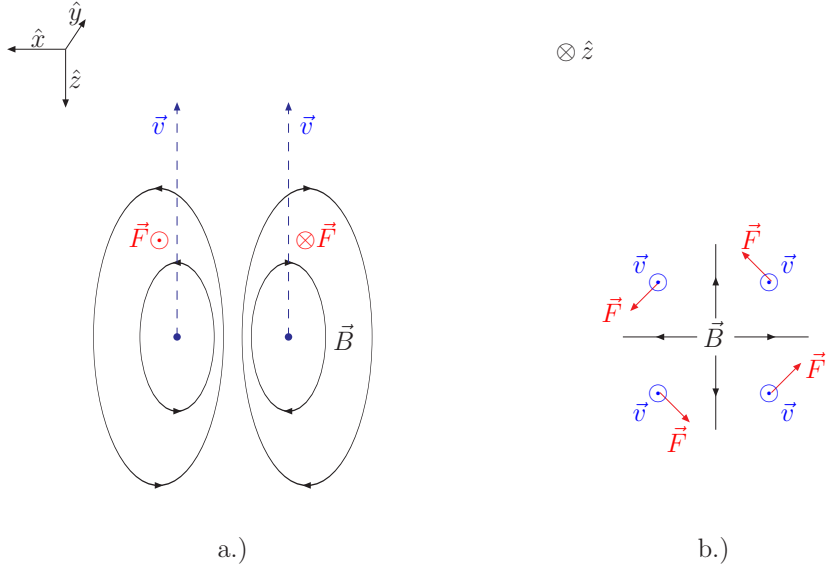}
\end{picture}
\caption{Illustration of the torque acting on the struck quark in the $-\hat{z}$ direction
through a color-magnetic dipole field caused by the spectators. a.) side view; b.) top view.
In this example the $\hat{z}$ component of the torque is negative as the quark leaves the
nucleon.
}
\label{fig:dipole}
\end{figure}
\acknowledgments
I would like to thank C. Lorc\'e and M. Wakamatsu for 
useful discussions. This work was supported by the DOE under grant number 
DE-FG03-95ER40965.

\appendix
\section{Gauge Invariance}
Although we discussed the difference between ${\cal L}^q$ and
$L^q$ in $A^+=0$ gauge in order to keep the equations simple, the
interpretation of their difference, as the change in orbital angular momentum, is manifestly gauge invariant. To see this, we first consider (for
simplicity, we only write down the expression in the abelian case --- in the
nonabelian case there are additional gauge links connecting the quark and gluon operators)
\begin{eqnarray}
{\cal L}_{+}^q  \label{eq:Lpm2}
= \frac{ \int d^3{\vec r}\langle PS | 
\bar{q}({\vec r}) \gamma^+\left[{\vec r}\times\left( i{\vec \partial}+ g {\cal A}_\perp ({\vec r})\right)\right]^zq({\vec r})|PS\rangle}
{\langle PS |PS\rangle}, \nonumber
\end{eqnarray}
in an arbitrary gauge, which also involves
a contribution from the gauge links to $x^-=\infty$
\begin{equation}
{\cal A}_\perp ({\vec r}) \equiv A_\perp({\vec r}_\perp,\infty)  - \int_{r^-}^\infty
dy^- \partial_\perp A^+({\vec r}_\perp,y^-) 
\label{eq:ALC}
\end{equation}
${\cal L}_+^q-L^q$ thus contains the matrix element of
$\bar{q}({\vec r}) \gamma^+\left[{\vec r}\times\left( g{\cal A}_\perp ({\vec r})-gA_\perp({\vec r})\right)\right]^zq({\vec r})$. Using 
\begin{equation}\hspace*{0.7cm}
{\cal A}_\perp ({\vec r})\!-\!A_\perp({\vec r}) \!=\! A_\perp({\vec r}_\perp,\infty)\!-\!A_\perp({\vec r}) \!-\! \int_{r^-}^\infty\!\!\!\!\!\!
dy^- \partial_\perp A^+({\vec r}_\perp,y^-) \!=\!\!
\int_{r^-}^\infty \!\!\!\!\!\!dy^- G^{+\perp}({\vec r}_\perp, r^-),
\end{equation}
which is identical to the terms entering Eq. (\ref{eq:torque}).


\begin{thebibliography}{0}

\bibitem{EMC} \BY{Ashman~J et al. (EMC)} \IN{Phys. Lett. B}{206}{1988}{364}; \IN{Nucl. Phys. B}{328}{1989}{1}

\bibitem{JiPRL} \BY{Ji~X} \IN{Phys. Rev. Lett.}{\bf 78}{1997}{610}

\bibitem{JM} \BY{Jaffe~R L \atque Manohar~A} \IN{Nucl. Phys. B}{B337}{1990}{509}

\bibitem{LCWF} \BY{Brodsky~S~J et al.} \IN{Nucl. Phys. B}{593}{2001}{311} 

\bibitem{Wakamatsu} \BY{Wakamatsu~M} \IN{Phys. Rev. D}{81}{2010}{114010}; \IN{Eur. Phys. J. A}{44}{2010}{297}; \IN{Phys. Rev. D}{85}{2012}{114039}

\bibitem{wigner} \BY{Belitsky~A.V., Ji~X. \atque Yuan~F.} 
\IN{Phys. Rev. D}{69}{2004}{074014}.

\bibitem{jifeng} \BY{Ji~X, Xiong~X \atque Yuan~F} 
\IN{Phys. Rev. Lett.}{109}{2012}{152005}

\bibitem{metz} \BY{Meissner~S, Metz~A \atque Schlegel~M}
\IN{JHEP}{0908}{2009}{056}

\bibitem{hatta} \BY{Hatta~Y} \IN{Phys. Rev. D}{84}{2011}{041701}; 
\IN{Phys. Lett. B}{708}{2012}{186}

\bibitem{lorce2} \BY{Lorc\'e~C} \IN{Phys. Lett. B}{719}{2013}{185}.

\bibitem{lorce} \BY{Lorc\'e~C \atque Pasquini~B} \IN{Phys. Rev. D}{84}{2011}{014015}; \BY{Lorce~C et al.} \IN{Phys. Rev. D}{85}{2012}{114006}


\bibitem{ams}  \BY{Tangerman~R~D \atque Mulders~P~J} \IN{Phys. Rev. D}{51}{1995}{3357}

\bibitem{RPP} \BY{Burkardt~M., Miller~A. \atque, Nowak~W.-D.}
\IN{Rept. Prog. Phys.}{73}{2007}{016201}.

\bibitem{Mauro} \BY{Anselmino~M. et al.} \IN{Eur. Phys. J. A}{39}{2009}{89}.

\bibitem{jifengTMD} \BY{Belitsky~A~V, Ji~X \atque Yuan~F}
\IN{Nucl. Phys. B}{656}{2003}{165}; \BY{Boer~D, Mulders~P~J \atque Pijlman~F} 
\IN{Nucl. Phys. B}{667}{2003}{201}; 

\bibitem{BC} \BY{Burkardt~M \atque BC~H} \IN{Phys. Rev. D}{79}{2009}{071501}

\bibitem{mb:force} \BY{Burkardt~M} {arXiv:08103589}

\bibitem{mb:torque} \BY{Burkardt~M} {arXiv:1205.2916}{}{}{}






























\end{thebibliography}
\end{document}